\documentclass[doublecol]{epl2}
\usepackage{amsmath,amssymb}
\usepackage{graphics}

\begin{document}

\title{Functional importance of noise in neuronal information processing}

\author{Daqing Guo\inst{1}, Matja{\v z} Perc\inst{2,3}, Tiejun Liu\inst{1}, Dezhong Yao\inst{1}}
\shortauthor{Guo et al.}
\institute{
\inst{1} The Clinical Hospital of Chengdu Brain Science Institute, MOE Key Lab for Neuroinformation, School of Life Science and Technology, University of Electronic Science and Technology of China, Chengdu 610054, China\\
\inst{2}Faculty of Natural Sciences and Mathematics, University of Maribor, Koro{\v s}ka cesta 160, SI-2000 Maribor, Slovenia\\
\inst{3}Complexity Science Hub Vienna, Josefst\"{a}dterstra{\ss}e 39, A-1080 Vienna, Austria}


\abstract{Noise is an inherent part of neuronal dynamics, and thus of the brain. It can be observed in neuronal activity at different spatiotemporal scales, including in neuronal membrane potentials, local field potentials, electroencephalography, and magnetoencephalography.  A central research topic in contemporary neuroscience is to elucidate the functional role of noise in neuronal information processing. Experimental studies have shown that a suitable level of noise may enhance the detection of weak neuronal signals by means of stochastic resonance. In response, theoretical research, based on the theory of stochastic processes, nonlinear dynamics, and statistical physics, has made great strides in elucidating the mechanism and the many benefits of stochastic resonance in neuronal systems. In this perspective, we review recent research dedicated to neuronal stochastic resonance in biophysical mathematical models. We also explore the regulation of neuronal stochastic resonance, and we outline important open questions and directions for future research. A deeper understanding of neuronal stochastic resonance may afford us new insights into the highly impressive information processing in the brain.}

\pacs{05.45.-a}{Nonlinear dynamics and chaos}
\pacs{87.19.L-}{Neuroscience}
\pacs{87.17.-d}{Cell processes}

\maketitle

\section{Introduction}
The human brain is composed of nearly 100 billion neurons that conduct signals via synapses~\cite{Kandel2012}. The massive number of neurons is believed to be the fundament behind the computational power of the brain, which integrates and generates neuronal information in the form of action potentials, or so-called spikes~\cite{Kandel2012, Gerstner2002}. Deeply understanding the mechanisms of signal processing amongst neurons promises to unlock the secrets behind the computational principles of the real brain.

Neuronal activity recorded in electrophysiological studies typically exhibits a certain level of stochastic fluctuations~\cite{destexhe2012, Stein2012}. This stochastic feature can be observed even when our brain is at rest, suggesting that biological neurons might operate in noisy environments.  To date, many types of noise sources have been identified in the brain, such as thermal noise~\cite{Stevens1972}, ion channel noise~\cite{Hille1992, Schmid2001, Strassberg1993, Linaro2011, Goldwyn2011}, synaptic release noise~\cite{Koch1999, Smetters1996, Branco2009, Guo2010, Uzuntarla2012, Guo2012}, and synaptic bombardment noise~\cite{Brunel2000, Brunel2001, Ho2000, Guo2012B}. Among them, the stochastic opening and closing of ion channels and massive amounts of synaptic bombardment are thought to be two main noise sources in neural systems. Because the firing dynamics of neurons can be largely impacted and disrupted by a stochastic drive, a natural question arises regarding whether neuronal noise always acts as a destructor in the brain. However, accumulating evidence has indicated that neuronal noise might promote signal processing of neurons under proper conditions~\cite{Stein2012, MaWJ2006,  Fanelli2017, McDonnell2011, McDonnell2009}.

Perhaps the most prominent noise-induced counterintuitive behavior discovered in the past 40 years is stochastic resonance (SR)~\cite{Benzi1982, Gammaitoni1998, McNamara1998, Berdichevsky1996, Gingl1995}. By definition, SR initially refers to a nonlinear system driven by a periodic signal showing the maximal information transmission at an intermediate noise level~\cite{Benzi1982, Gammaitoni1998, McNamara1998, Berdichevsky1996, Gingl1995}. Neurons in the brain are not only driven by stochastic fluctuations from various noise sources but also receive rhythmic signals due to neuronal oscillations~\cite{Kandel2012, Gerstner2002}. Furthermore, biological neurons exhibit different types of firing excitability and show highly nonlinear responses to external stimuli~\cite{Izhikevich2007}. These features enable the SR to serve as an underlying dynamical mechanism that real neurons utilize to improve information processing, an idea supported by increasing experimental data. For instance, Douglass et al. showed that appreciable level of external noise can enhance the information transfer in mechanoreceptive sensory neurons of crayfish, indicating the occurrence of SR in neural systems~\cite{Douglass1993}. Similar neuronal SR phenomena have been widely observed in sensory systems of other species, such as crickets~\cite{Levin1996}, rats~\cite{Collins1996} and paddlefish~\cite{Russell1999}. Later investigations have also demonstrated SR in neurons from the mammalian brain~\cite{Gluckman1996, Gluckman1998}.

Recent studies using computational modeling have provided more evidence of the benefits of SR occurring in neural systems~\cite{McDonnell2011, McDonnell2009}. Specifically, it has been demonstrated that different types of SR may arise in neural systems and can be regulated by several intrinsic and external properties. In this mini review, we briefly summarize and discuss computational results on the functional roles of noise in enhancing neuronal information processing via SR-related mechanisms and propose several urgent questions in this research field.

\section{Models and measurements for neuronal SR studies}
\begin{figure}[!t]
	\includegraphics[width=8.8cm]{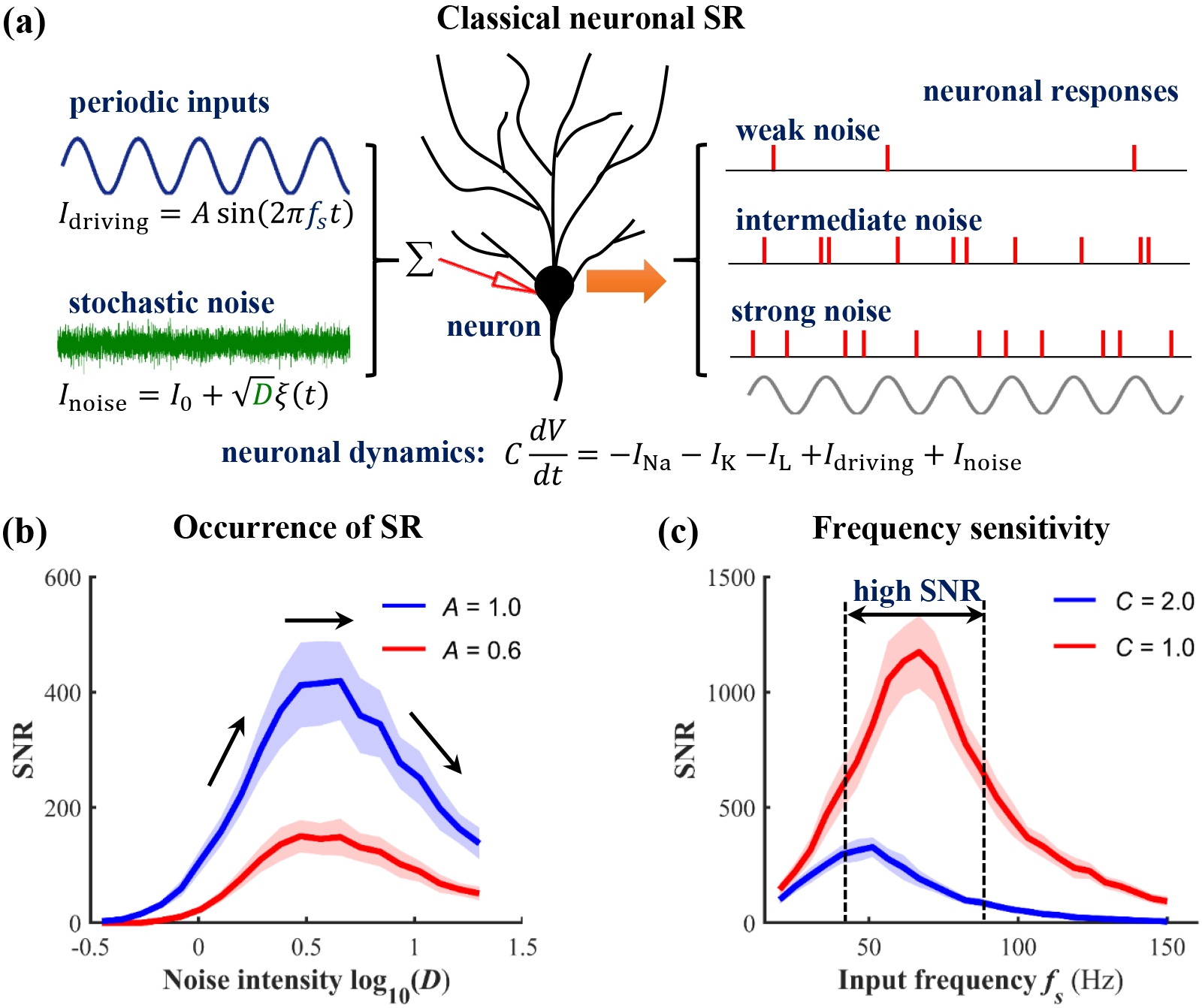}
	\caption{(Color online)
		Noise-induced SR in neural systems. (a) Framework for classical neuronal SR studies. A neuron is simultaneously driven by periodic inputs and stochastic noise. The neuron optimally responds to the periodic signal at an intermediate noise level. (b) Dependence of the SNR on the noise intensity $D$. A bell-shaped SNR curve is seen by varying $D$, indicating the occurrence of SR. At the same noise level, a relatively stronger subthreshold periodic driving signal evokes a higher SNR value. (c) Dependence of SNR on the input frequency $f_s$. For different membrane capacitances, the neuron exhibits distinct frequency sensitivity ranges. Data depicted in (b) and (c) are based on simulations using the HH model.}
	\label{fig:1}
\end{figure}

Classical neuronal SR studies focus on a neuron driven by both a subthreshold periodic signal and stochastic noise (Fig.~\ref{fig:1}(a))~\cite{Longtin1993, Lee1999, Bulsara1991}. As a paradigmatic model, the Hodgkin-Huxley (HH) neuron has served as the preferred model for simulating neuronal dynamics~\cite{Gerstner2002, Hodgkin1952}. In classical neuronal SR studies, the current balance equation of the HH neuron can be written as:
\begin{equation}
\begin{split}
C\frac{dV}{dt}=-I_{\text{Na}}-I_{\text{K}}-I_{\text{L}}+I_{\text{driving}}+I_{\text{noise}},
\end{split}
\label{eq:1}
\end{equation}
where $V$ represents the membrane potential and $C$ is the membrane capacitance per unit area. The sodium, potassium, and leakage currents through the membrane are modeled as $I_{\text{Na}}=G_{\text{Na}}m^3h(V-E_{\text{Na}})$, $I_{\text{K}}=G_{\text{K}}n^4(V-E_{\text{K}})$ and $I_{\text{L}}=G_{\text{L}}(V-E_{\text{L}})$, respectively. The model parameters $G_{\text{Na}}$, $G_{\text{K}}$ and $G_{\text{L}}$ are the maximal sodium, potassium, and leakage conductances per unit area, and $E_{\text{Na}}$, $E_{\text{K}}$ and $E_{\text{L}}$ denote the corresponding reversal potentials. Three gating variables obey the following Langevin equation~\cite{Gerstner2002, Hodgkin1952}:
\begin{equation}
\begin{split}
\frac{dx}{dt}=\alpha_x(V)(1-x)-\beta_x(V)x,
\end{split}
\label{eq:2}
\end{equation}
where $x=m, h, n$, and $\alpha_x(V)$ and $\beta_x(V)$ are six voltage-dependent rate functions given in~\cite{Gerstner2002, Hodgkin1952}. The subthreshold periodic driving can be mimicked by a sinusoidal signal, $I_{\text{driving}}=A\sin(2\pi f_s t)$, with $A$ and $f_s$ representing signal amplitude and forcing frequency, respectively. The noise current is typically modeled as $I_{\text{noise}}=I_0 + \sqrt{D}\xi(t)$, where $I_0$ denotes the bias current, $\xi(t)$ is the Gaussian white noise with zero mean and unit variance, and $D$ is a parameter controlling the noise intensity. To a certain extent, this type of noise current can be used to simulate membrane fluctuations due to massive synaptic bombardment~\cite{Brunel2000, Brunel2001, Guo2012B}. Note that, as another important type of neuronal noise, the ion channel noise can be modeled by adding stochastic terms in the Langevin equations for gating variables~\cite{Schmid2001, Linaro2011, Goldwyn2011}:
\begin{equation}
\begin{split}
\frac{dx}{dt}=\alpha_x(V)(1-x)-\beta_x(V)x + \xi_x(t),
\end{split}
\label{eq:3}
\end{equation}
where $\xi_x(t)$ ($x=m, h, n$ ) are independent Gaussian white noises with zero mean. For different gating variables, the noise autocorrelation functions depend on the stochastic membrane potential and the total number of ion channels controlled by both the densities of ion channels and the area of the membrane patch, which has been described in previous modeling studies in detail~\cite{Schmid2001, Linaro2011, Goldwyn2011}.

The information transfer capability for neurons driven by external periodic inputs can be quantified by several measurements. Similar to SR in other dynamical systems, the most frequently used measure is the signal-to-noise ratio (SNR) at the input frequency $f_s$~\cite{McDonnell2009}. To compute the SNR, the power spectral density of the spike train is estimated using a fast Fourier transform. Mathematically, the SNR is defined as~\cite{Lee1999, Guo2012B}:
\begin{equation}
\begin{split}
\text{SNR}=\frac{S(f_s)-N(f_s)}{N(f_s)},
\end{split}
\label{eq:4}
\end{equation}
where $S(f_s)$ is the power at the input frequency $f_s$, and $N(f_s)$ is the mean background power at nearby frequencies. As a signature of neuronal SR, it can be seen that the SNR curve exhibits a bell-shaped profile with increasing noise intensity (Fig.~\ref{fig:1}(b)). On the other hand, the second measure widely employed in neuronal SR studies is the Fourier coefficient~\cite{Ozer2009A, Ullner2003}. This measurement has been confirmed to be proportional to the square of the spectral power amplification and can be calculated as~\cite{Holden1993}:
\begin{equation}
\begin{split}
Q=\sqrt{Q_{\text{sin}}^2 + Q_{\text{cos}}^2},
\end{split}
\label{eq:4}
\end{equation}
where $Q_{\text{sin}}=\frac{1}{nT}\int_{0}^{nT}2V(t) \sin(2\pi t/T)dt$ and $Q_{\text{cos}}=\frac{1}{nT}\int_{0}^{nT}2V(t) \cos(2\pi t/T)dt$. Here, $T=1/f_s$ is the period of the external driving signal, and $n$ is a positive integer related to the simulation time. Similar to SNR, a larger $Q$ value implies a better information transfer capability. When SR behavior occurs in neural systems, a peak value can be seen in the $Q$ curve as a function of noise intensity. Note that several other measurements, such as power norms~\cite{Collins1995A, Collins1995B, Collins1996B} and information-based measures~\cite{Stocks2001, Hoch2003, Durrant2011}, can also be used to characterize the performance of SR.

\section{Stochastic resonance in neural systems}
Computational models offer an efficient approach to investigate neuronal SR. Following experimental observations, recent theoretical explorations have suggested that distinct types of SR may be evoked in neural systems under different circumstances. In this section, we briefly summarize several typical neuronal SR behaviors observed in modeling studies.

As described above, the classical neuronal SR framework mainly considers a neuron or a population of neurons driven by both stochastic noise and the subthreshold periodic force with a single frequency component~\cite{Longtin1993, Lee1999, Bulsara1991}. By computational modeling, the classical SR has been demonstrated to appear in spiking model neurons with either threshold-spiking or resonance-spiking mechanisms~\cite{Lee1999, Guo2012B, Gammaitoni1998, Plesser1999}. These two generation mechanisms of spikes involve firing dynamics for most biological neurons and correspond to class I and II neuronal excitability~\cite{Izhikevich2007}, respectively. As shown in Fig.~\ref{fig:1}(b), the performance of classical neuronal SR is impacted by the amplitude of the external periodic signal. A relatively stronger subthreshold periodic driving signal tends to evoke a higher peak SNR. Furthermore, this amplitude-based neuronal SR is also modulated by the input frequency of the external driving signal (Fig.~\ref{fig:1}(c)). It has been found that neurons show optimal responses to the external periodic signal at a special frequency range, implying the existence of frequency sensitivity. Similar neuronal frequency sensitivity has been widely reported in past experimental and computational studies~\cite{Guo2012B, Liu1999A, Yu2001, Guo2009}, which can be attributed to the cooperation of the intrinsic neuronal oscillations and the periodic input signals. This observation is of particular interest because oscillatory signals in the brain cover multiple frequency bands~\cite{Kandel2012}, and neurons with different intrinsic spiking dynamics might preferentially process weak neural information with distinct frequency bands via SR mechanism (see Fig.~\ref{fig:1}(c)).

\begin{figure}[!t]
	\includegraphics[width=8.8cm]{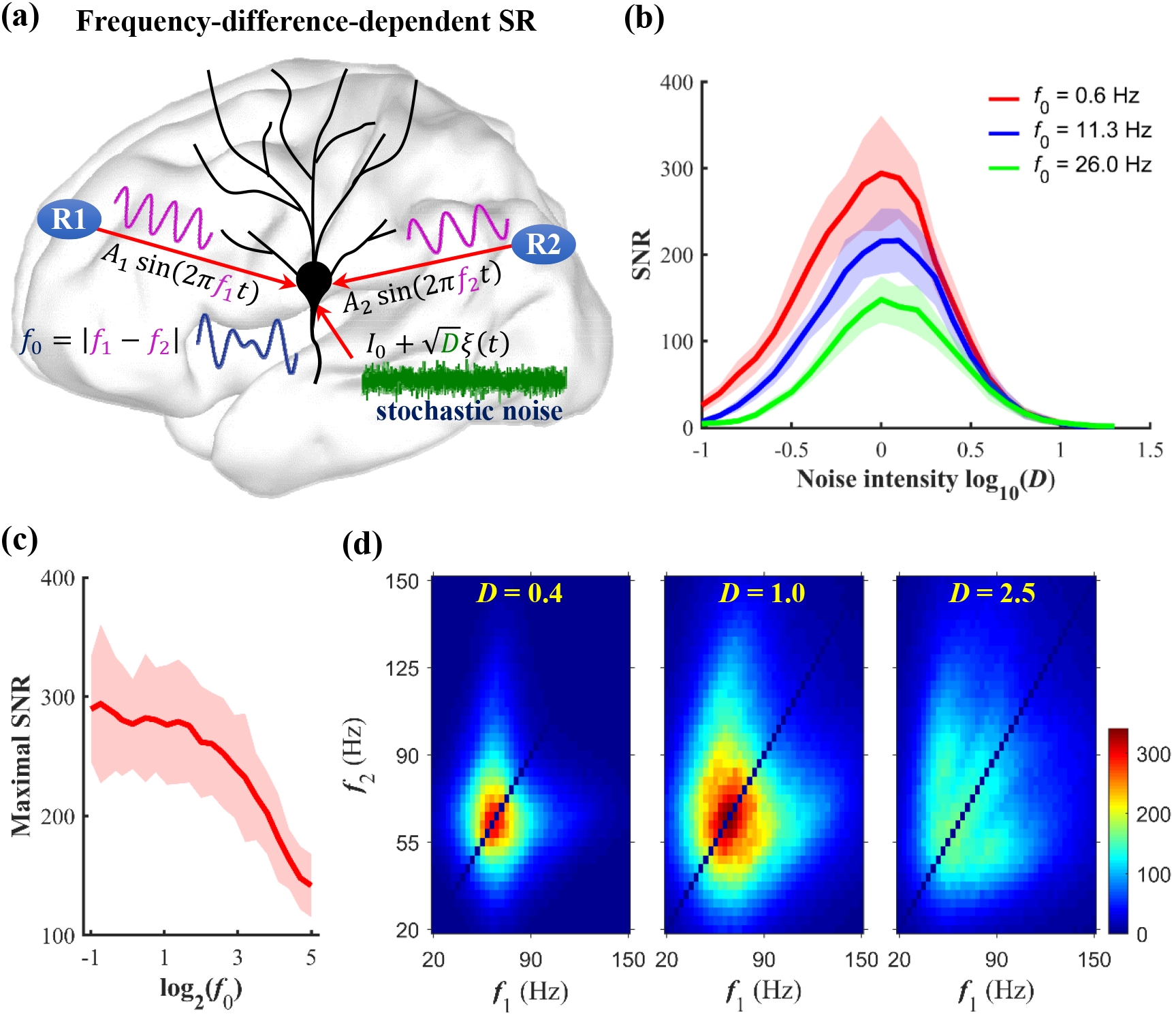}
	\caption{(Color online)
		Noise-induced frequency-difference-dependent SR (FDDSR) in neural systems. (a) Schematic presentation of FDDSR studies. A neuron is driven by stochastic noise and two periodic inputs with frequencies $f_1$ and $f_2$. (b) Dependence of the SNR on noise intensity $D$ at different beat frequencies $f_0=|f_1-f_2 |$. (c) Maximal SNR versus $f_0$. The neuron shows a better performance at a relatively lower beat frequency. (d) Dependence of the SNR on driving frequencies $f_1$ and $f_2$. The neuron exhibits a wider frequency-sensitivity range at an intermediate noise level. The data presented here are adapted from our previous study~\cite{Guo2017}.}
	\label{fig:2}
\end{figure}

On the other hand, biological neurons may simultaneously receive more complicated oscillatory signals from different brain regions with mixed-frequency features. Thus, a naturally arising question is whether similar SR phenomena can also be discovered in neural systems with multiple periodic input components. As pioneers in this field, Chialvo et al. confirmed that a single neuron driven by mixed periodic signals with harmonic frequencies of a fundamental frequency showed the maximal response to the fundamental frequency at an intermediate noise level~\cite{Chialvo2002, Chialvo2003, Chialvo2012}. They labeled this seemingly SR-like effect as the ``ghost SR'' (GSR), implying that it appears at the fundamental frequency missing in the input signals~\cite{Chialvo2002, Chialvo2003, Chialvo2012}. Recently, we extended the theory of ghost SR by investigating the response of neural systems to two subthreshold periodic signals with an arbitrary difference in frequency (Fig.~\ref{fig:2}(a))~\cite{Guo2017}. Through computational modeling, we demonstrated that SR might occur at the beat frequency in neural systems at both the single-neuron and population levels (Fig.~\ref{fig:2}(b)). Similarly to the classical SR in neural systems~\cite{Liu1999A, Yu2001, Guo2009}, the performance of this frequency-difference-dependent SR exhibits a frequency sensitivity feature, and a smaller beat frequency corresponds to a stronger SNR at the optimal noise level (Figs.~\ref{fig:2}(c) and~\ref{fig:2}(d)). Furthermore, our simulations indicated that the population response of neural ensembles is more efficient than that of a single neuron to detect neural information carried by the superposition of multiple periodic signals at the beat frequency. These results highlight the functional roles of stochastic noise in enhancing the signal transduction for beat-frequency-related neural information and may have important biological applications.

\begin{figure}[!t]
	\includegraphics[width=8.8cm]{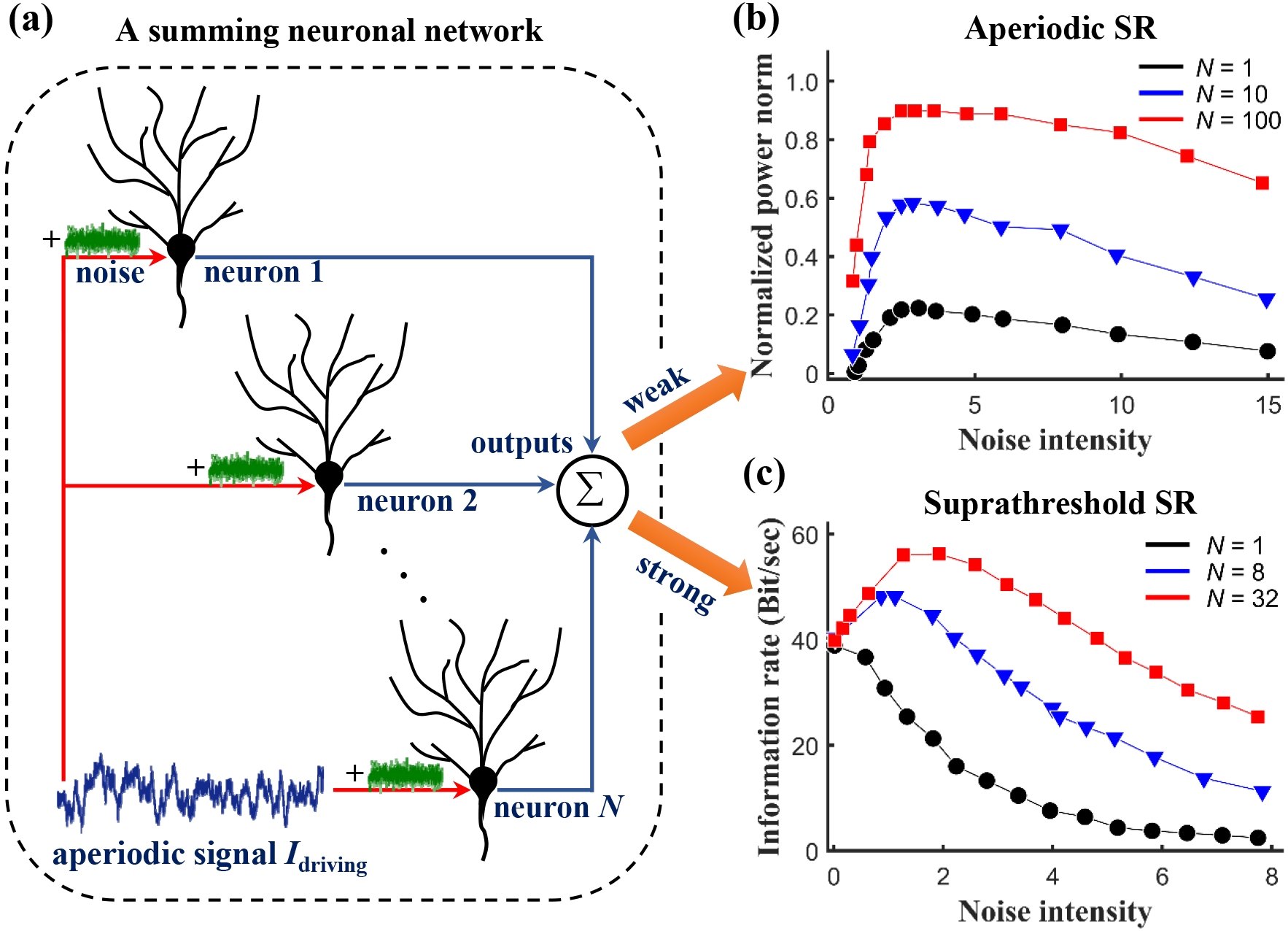}
	\caption{(Color online)
		Aperiodic SR and suprathreshold SR in neural systems. (a) Schematic presentation of a summing network consisting of $N$ neurons. Each neuron is driven by a common aperiodic signal and independent stochastic noise. (b) Dependence of the normalized power norm on noise intensity. (c) Dependence of the information rate on noise intensity. Aperiodic SR appears for a weak common driving signal, whereas suprathreshold SR can be observed at the population level for a strong common driving signal. The data presented here are adapted from previous studies~\cite{Collins1995A, Hoch2003}.}
	\label{fig:3}
\end{figure}

Although neuronal SR studies have commonly assumed weak periodic external forces, it should be noted that the SR-type behaviors in neural systems are not limited to signals with periodic components and signals below the threshold level (Fig.~\ref{fig:3}(a)). Using a variety of spiking model neurons, Collins et al. developed a theory that the SR effect can be detected with the correlation-based power norm when a neuron is subjected to weak aperiodic signals and white noise (Fig.~\ref{fig:3}(b))~\cite{Collins1995A, Collins1995B, Collins1996B}. Inspired by the aperiodic property of the external driving signal, this phenomenon was designated as aperiodic SR~\cite{Collins1995A, Collins1995B, Collins1996B}. Compared with a single neuron, the detection capability of a weak signal via aperiodic SR can be further improved in a summing network consisting of a population of neurons. This theoretical prediction has been demonstrated by well-designed biological experiments, showing that aperiodic SR indeed exists in mammalian sensory neurons~\cite{Collins1996B}. Using information-based measurements, numerical simulations have shown that SR-type behavior can also be observed in a summing neuronal network even when the external driving signal is sufficiently strong and above the threshold of neurons (Fig.~\ref{fig:3}(c))~\cite{Stocks2001, Hoch2003, Durrant2011}. However, the bell-shaped measurement curve disappears for a single neuron (see $N=1$ in Fig.~\ref{fig:3}(c)), indicating that the suprathreshold neuronal SR may be only observed at the population level. Further experiments need to be designed and performed to verify whether suprathreshold SR indeed exists in the real brain.

We emphasize that many other types of SR-related behaviors may also occur in neural systems, and several of them are described as follows. First, although previous modeling studies have shown that neurons operating as coincidence detectors (class III excitability) may not show classical SR behaviors, this type of neuron has been identified as sensitive to changes in the stimulus~\cite{Gai2010}. In the presence of noise, this dynamic feature causes class III neurons to exhibit slope-based SR behavior, a phenomenon supported by both experimental and computational evidence~\cite{Gai2010, Schmerl2013}. Second, even without an external driving signal, neurons perturbed solely by noise may also display so-called internal or autonomous SR behavior~\cite{Longtin1997}. This phenomenon is highly associated with coherence resonance in excitable systems~\cite{Pikovsky1997}, and a noise-induced peak can be seen clearly in the power spectrum. As the noise intensity is increased, the SNR value first drops and then rises, and the internal SR appears at an intermediate level of neuronal noise. This phenomenon has not only been observed in theoretical studies but has also been confirmed in recent biological experiments~\cite{Manjarrez2002}. Third, a similar enhancement effect of stochastic noise can also be induced by a high-frequency signal via vibrational resonance (VR)~\cite{Landa2000}. As an important variant of SR, the phenomenon of VR has been observed in many neural systems, in which the neuronal response to a low-frequency external driving signal is improved at the optimal amplitude of the high-frequency signal~\cite{YuHT2011, SunJB2013}. This finding is of biological importance because high-frequency neural oscillations have been widely observed in the brain and have been linked to many higher brain functions~\cite{Bragin1999}. Finally, many nonlinear systems stimulated by subthreshold periodic signals have been found to be enhanced at more than one noise level~\cite{Vilar1997, Zeng2011}. Note that this specific type of SR is referred to as stochastic multi-resonance or multiple SR, which has also been widely reported in recent neural modeling studies~\cite{LiHY2018, Pinamonti2012, Mejias2011}.

Overall, these observations provide both the computational evidence and theoretical basis for the occurrence of SR in neural systems. After long-term evolution, our brain might have the ability to exploit SR by optimizing stochastic noise from different sources to facilitate neuronal information processing.

\section{Regulation of SR in neural systems} What are the regulatory mechanisms of neuronal SR in the brain? Theoretically, there are several underlying intrinsic and external biological factors that can achieve this function, and we discuss this issue from different levels in the following section.

\begin{figure}[!t]
	\includegraphics[width=8.8cm]{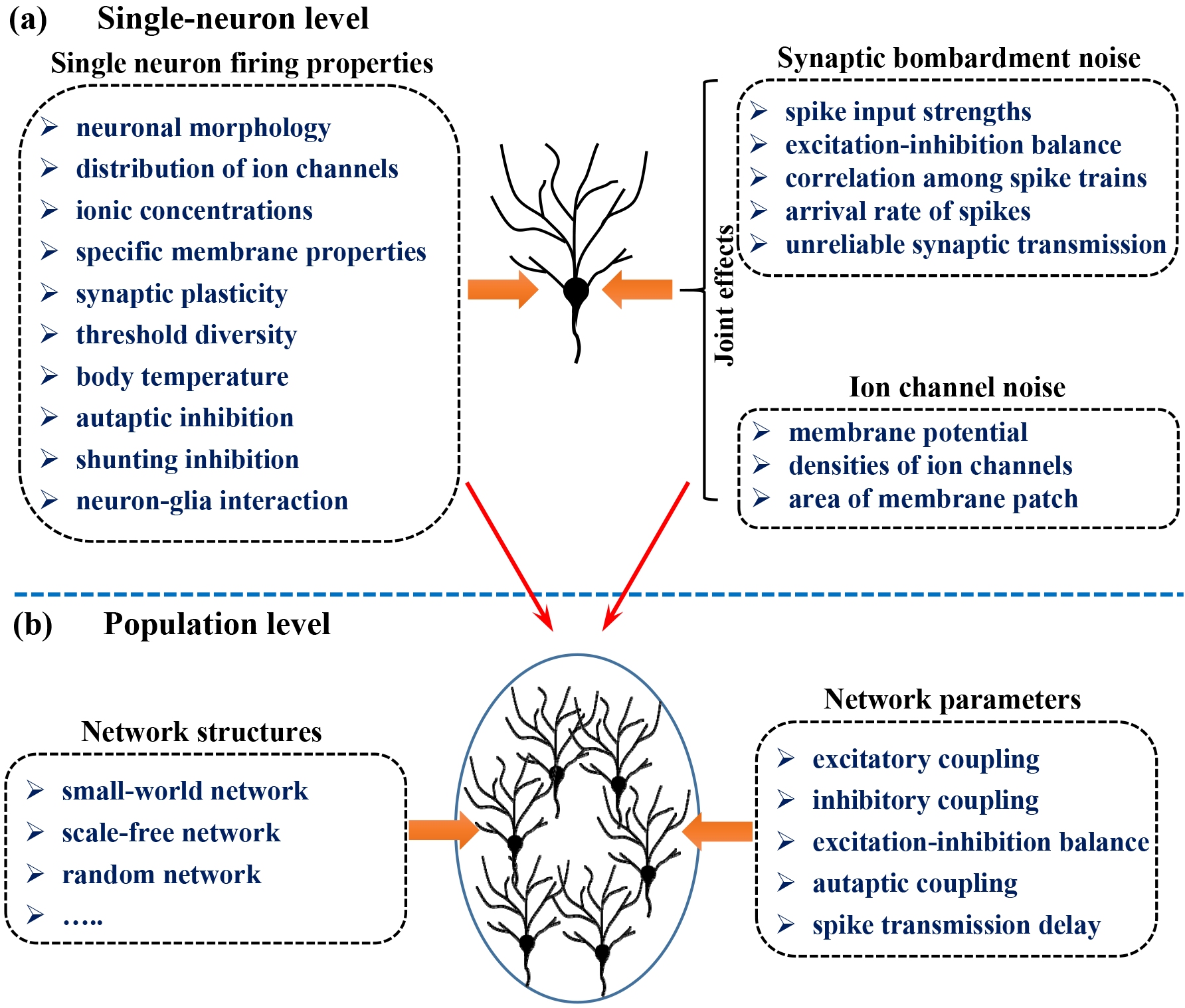}
	\caption{(Color online)
		Regulation of neuronal SR at different levels. (a) At the single-neuron level, neuronal SR is mainly impacted by single neuron firing properties and joint effects of different types of neuronal noise. (b) At the population level, neurons respond to external driving signals together, and the performance of neuronal SR is both influenced by regulatory factors at the signal-neuron level and modulated by the network structure and parameters.}
	\label{fig:4}
\end{figure}

At the single-neuron level, both the weak signal detection capability and the frequency sensitivity range of neurons highly depend on the single neuron firing properties (Fig.~\ref{fig:4}(a)). Experimental studies have established that intrinsic differences in neuronal morphology, the distribution of ion channels, ionic concentrations, specific membrane properties and threshold diversity might result in the distinctive firing properties of neurons~\cite{Connors1996, Arhem2006, Zibman2011, Zeng2013, Cervera2013}. In addition, several other external factors, such as body temperature~\cite{YuYG2012}, autaptic and shunting inhibition~\cite{Prescott2006, Guo2016A}, short-term plasticity~\cite{Mejias2011} and neuron-glia interactions~\cite{Wang2012A}, may also dramatically change neuronal firing properties. We predict that these intrinsic and external factors may play functional roles in regulating neuronal SR. On the other hand, real neurons are driven by different types of stochastic noise in the brain. In theory, neuronal noise stemming from different sources should be modulated by distinct biological mechanisms~\cite{destexhe2012}. As mentioned above, synaptic bombardment noise and ion channel noise are two main types of noise in neural systems, and their joint effect may dominate the stochastic dynamics of neuronal firing (Fig.~\ref{fig:4}(a)). Using the mean-field theory, previous studies have shown that synaptic bombardment noise is determined largely by the strengths of excitatory and inhibitory inputs, the balance between excitation and inhibition, the correlation among input spike trains and the effective mean arrival rate of spikes~\cite{Guo2012B, Durrant2011, Kreuz2006}; furthermore, unreliable synaptic transmission might be an important mechanism regulating this type of neuronal noise~\cite{Guo2012B}. Ion channel noise has been found to be modulated by the membrane potential of neurons, the densities of ion channels and the area of the membrane patch~\cite{Schmid2001, Linaro2011, Goldwyn2011}. In the real brain, we postulate that these two types of neuronal noise may work together and be jointly responsible for evoking SR at the single-neuron level.

Compared with a single neuron, a population of neurons may cooperate with each other and collectively respond to a weak external driving signal. At an optimal noise intensity, many studies have shown that the maximal response of neuronal ensembles to a weak driving signal is relatively stronger than that at the single-neuron level~\cite{Collins1995A, Collins1995B, Collins1996B,Stocks2001, Hoch2003, Durrant2011, Guo2017}. These observations provide evidence that collective neuronal firing might be more efficient in detecting neuronal information carried by weak driving signals. However, the performance of population neuronal SR is not only impacted by the abovementioned intrinsic and external regulatory factors occurring at the signal-neuron level but can also be significantly modulated by both the network structure and network parameters (Fig.~\ref{fig:4}(b)), which are systematically summarized as follows.

First, intensive statistical analysis of the neuronal connectome at the micro-scale level has revealed that the wiring diagram of neurons exhibits typical small-world and scale-free features depending on different conditions~\cite{Varshney2011}. Indeed, there is accumulating computational evidence that the performance of neuronal SR can be amplified via fine-tuning of the small-world or scale-free network structures~\cite{Ozer2009A, Perc2007, Yilmaz2013, Wu2009}. More importantly, using a spatially embedded network model, an optimal topology between the heterogeneous scale-free network and the strong random geometric network has also been determined to optimize the performance of neuronal SR, revealing that a suitable number of hubs and with that an optimal ratio between short- and long-range connections is critical for enhancing SR in neural systems~\cite{Gosak2011}.

Second, neurons communicate with each other through synapses, and different types of synaptic couplings may contribute distinct effects to neuronal SR (Fig.~\ref{fig:4}(b)). The most common type of synapse in the brain is the chemical synapse, which can be further divided into excitatory and inhibitory synapses~\cite{Kandel2012}. By establishing neuronal networks coupled by chemical synapses, several studies have demonstrated that a network with strong excitation or inhibition dramatically deteriorates the weak signal detection capability, and a suitable excitation-inhibition balance in the network can improve the performance of neuronal SR-related behaviors~\cite{Guo2017, Wang2016A}. Furthermore, electrical synapses have also been demonstrated to strongly impact the dynamics of neural systems. Using a small-world network composed of Rulkov map-based neurons, Perc reported that the performance of neuronal SR induced by a pacemaker highly depends on the electrical coupling strength, and the small-world network property is able to enhance the SR only for intermediate levels of coupling strength~\cite{Perc2007}. Note that a similar effect of electrical synapses has also been found in FitzHugh-Nagumo (FHN) neurons connected in an array~\cite{Kanamaru2001}, but such enhancement seems not to be reproduced in a small-world neuronal network consisting of HH neurons~\cite{Ozer2009A}. In that study, increasing the electrical coupling strength tends to decrease the maximal neuronal response to weak periodic driving signals~\cite{Ozer2009A}. Interestingly, several studies have also compared the effects of chemical and electrical synapses on the enhancement of signal propagation via SR-related mechanisms in neuronal networks with different structures~\cite{SunJB2013, LiXM2007, Yilmaz2013}. Most relevant studies have indicated that excitatory synaptic coupling might be more efficient than electrical synapses in weak signal detection~\cite{SunJB2013, LiXM2007}. In addition to normal ``feedforward'' synapses, neurons might also be self-innervated via feedback connections referred to as autapses~\cite{Guo2016A}. It was recently demonstrated that the performance of pacemaker-induced SR can be improved in a scale-free network composed of HH neurons at an optimal autaptic coupling strength~\cite{Yilmaz2016}.

Lastly, spike transmission delay due to both the finite propagation speed and the time lapse occurring by dendritic and synaptic integrations is unavoidable in neural systems~\cite{Kandel2012, Gerstner2002}. As an intrinsic property of neuronal information processing, the transmission delay of spikes may play a vital role in regulating neuronal SR (Fig.~\ref{fig:4}(b)). In particular, several modeling studies have shown that the spike transmission delay is critical for triggering multiple SR, and such behavior appears at every integer multiple of the period of the external driving signal~\cite{Wang2009, Hao2011}. By introducing a pacemaker with an autapse, it has also been reported that multiple SR can be evoked by matched autaptic transmission delay in neuronal networks~\cite{Yilmaz2016}. These results, however, do not suggest that the transmission delay is a necessary condition for multiple SR in neural systems. For a sufficiently long period of subthreshold driving signals, neural systems may exhibit typical multiple SR even in the absence of spike transmission delay~\cite{LiHY2018}. Additionally, the transmission delay may also modulate the performance of neuronal SR. Such modulation of neuronal SR shows high sensitivity to stochastic noise~\cite{SunXJ2018}. At an optimal noise level, introducing the spike transmission delay into the network may reduce the capability of weak signal detection, thus deteriorating the performance of neuronal SR. When stochastic noise is not at the optimal level, a suitable tuning of transmission delay is able to assist the ability of noise to detect subthreshold driving signals.

\section{Conclusions and open questions} Random fluctuations in brain activity are attributed to diverse sources of neuronal noise and have been widely observed in experimental recordings at different spatiotemporal scales~\cite{destexhe2012, Stein2012}. At the micro-scale level, neuronal membrane potentials also display the feature of voltage fluctuations in millisecond temporal resolution, strongly implying that neurons may operate in noisy environments. Recent studies have shown that a suitable level of neuronal noise may not destroy but can guarantee efficient signal processing in the brain~\cite{Stein2012, MaWJ2006, Fanelli2017, McDonnell2011, McDonnell2009}. In this context, SR is a leading candidate mechanism for enhancing the capability of weak signal detection in neural systems. Over the last several decades, remarkable progress has been made in this field using computational approaches\cite{McDonnell2011, McDonnell2009}. In this mini review, we systematically summarized different types of SR-related behaviors observed in modeling studies and also discussed several possible regulation mechanisms of neuronal SR in the brain.

Although neuronal SR has been widely investigated in recent years, there are still several open questions that deserve further exploration. First, an increasing number of modeling studies have documented different types of SR in neural systems, but it is not established whether some of them can coexist in the same system. In addition, it remains unknown whether all types of neuronal SR discovered in modeling studies can be observed in experiments. Further studies combining both computational and experimental approaches are needed to examine these two issues. On the other hand, chemical and electrical synapses coexist within the mammalian brain~\cite{Kandel2012, Gerstner2002}. However, previous modeling studies on neuronal SR have seldom included them together in the same network with an appropriate ratio matching experimentally reported data. Accordingly, it is still unclear whether these two types of synapses may perform combination and complementary roles in evoking and regulating neuronal SR, a question that can be further tested in future studies.

Looking forward, we must be cognizant that SR is only one possible mechanism by which neurons may exploit noise to facilitate signal processing in the brain. Other noise-induced behaviors, such as stochastic synchronization~\cite{Galan2006}, inverse stochastic resonance~\cite{Tuckwell2009, Guo2011, Uzuntarla2017, Yamakou2017},  phase resetting  of collective rhythm~\cite{Levnajic2010, Levnajic2011} and neural avalanches~\cite{Scarpetta2013}, may also play critical roles in high-efficiency neuronal information processing. A comprehensive theoretical framework that incorporates all these noise-enhanced phenomena will surely be rewarding in future modeling studies.

\section{Acknowledgments} We sincerely thank Dr. Dong-ping Yang for his kind help in proofreading our manuscript. This research was supported by the National Natural Science Foundation of China (Grant Nos.~31771149, 61527815, 81571770, 81771925 and 81371636), the Project of Science and Technology Department of Sichuan Province (Grant Nos.~2017HH0001 and 2018HH0003), and the Slovenian Research Agency (Grant Nos.~J1-7009, J4-9302, J1-9112 and P1-0403).


\end{document}